\documentclass[11pt]{article}
\pdfoutput=1
\usepackage{dcolumn}% Align table columns on decimal point
\usepackage{bm}% bold math
\usepackage{graphicx}% Include figure files
\usepackage{amssymb,amsmath}
\usepackage{ulem}
\usepackage{multirow}
\usepackage{cite,color,url}
\usepackage[colorlinks=true
,urlcolor=blue
,anchorcolor=blue
,citecolor=blue
,filecolor=blue
,linkcolor=blue
,menucolor=blue
,linktocpage=true
,pdfproducer=medialab
,pdfa=true
]{hyperref}

\usepackage{slashed}
\usepackage{epsfig,psfrag,rotating,soul}
\usepackage{rotfloat}
\usepackage[font={small}]{caption}
\usepackage{colortbl}

\usepackage{tikz} 
\usetikzlibrary{chains,shapes,fit,calc}
\usetikzlibrary{positioning,arrows}

\usepackage{geometry}
\usepackage{tikz-feynman}
\tikzfeynmanset{compat=1.0.0}

\usetikzlibrary{patterns,decorations.pathreplacing}
\usetikzlibrary{decorations.pathmorphing}
\usetikzlibrary{decorations.markings}
\usetikzlibrary{arrows,shapes}
\usetikzlibrary{shapes.misc}
\tikzset{
  gluon/.style={decorate, draw=black,
    decoration={coil,amplitude=4pt, segment length=4pt,aspect=0.7}} 
}
\tikzset{
  photon/.style={decorate, decoration={snake}},
}

\evensidemargin \oddsidemargin
\allowdisplaybreaks

\def\hc{\text{h.c.}}
\def\BR{\text{BR}}

\def\z2{$\mathbb{Z}_2$}

\definecolor{mygray}{gray}{0.85} 
\definecolor{myblue}{cmyk}{0.65, 0.37, 0.0, 0.19}

\definecolor{avblue}{rgb}{0.0, 0.0, 0.8}

\definecolor{asparagus}{rgb}{0.53, 0.66, 0.42}

\begin{document}
\thispagestyle{empty}

\def\thefootnote{\fnsymbol{footnote}}

\begin{flushright}
  IFIC/23-20
\end{flushright}
	
\vspace*{1cm}

\begin{center}
  
  \begin{Large}
    \textbf{\textsc{A Scotogenic explanation for the 95 GeV excesses}}
  \end{Large}
  
  \vspace{1cm}
  
  {\sc
    Pablo Escribano$^{a}$%
    \footnote{{\tt \href{mailto:pablo.escribano@ific.uv.es}{pablo.escribano@ific.uv.es}}}%
    , V\'ictor Mart\'in Lozano$^{a,b}$%
    \footnote{{\tt \href{mailto:victor.lozano@ific.uv.es}{victor.lozano@ific.uv.es}}}%
    , Avelino Vicente$^{a,b}$%
    \footnote{{\tt \href{mailto:avelino.vicente@ific.uv.es}{avelino.vicente@ific.uv.es}}}%
    
  }
  
  \vspace*{.7cm}
  
  {\sl
  (a) Instituto de F\'{i}sica Corpuscular, CSIC-Universitat de Val\`{e}ncia, 46980 Paterna, Spain \\
  
  (b) Departament de F\'{\i}sica Te\`{o}rica, Universitat de Val\`{e}ncia, 46100 Burjassot, Spain
  }
		
\end{center}

\vspace{0.1cm}

\begin{abstract}
  \noindent
Several hints of the presence of a new state at about $95$ GeV have
been observed recently. The CMS and ATLAS collaborations have reported
excesses in the diphoton channel at about this diphoton invariant mass
with local statistical significances of $2.9 \, \sigma$ and $1.7 \,
\sigma$, respectively. Furthermore, a $2 \, \sigma$ excess in the $b
\bar{b}$ final state was also observed at LEP, again pointing at a
similar mass value. We interpret these intriguing hints of new physics
in a variant of the Scotogenic model, an economical scenario that
induces Majorana neutrino masses at the loop level and includes a
viable dark matter candidate. We show that our model can naturally
explain the 95 GeV excesses while respecting all experimental
constraints and discuss other phenomenological predictions of our
scenario.
\end{abstract}

\def\thefootnote{\arabic{footnote}}
\setcounter{page}{0}
\setcounter{footnote}{0}

\newpage

\section{Introduction}
\label{sec:intro}

The Standard Model (SM) of particle physics has been an incredibly
successful framework for understanding the fundamental particles and
forces that make up our Universe. However, it faces significant
challenges when it comes to explaining two crucial phenomena: neutrino
masses and dark matter (DM). Neutrinos were long thought to be
massless, as suggested by the original formulation of the SM. However,
experimental evidence has now firmly established that neutrinos do
have masses, albeit very small ones. Similarly, the existence of DM,
which is inferred from its gravitational effects on visible matter,
poses another major challenge and various theoretical extensions of
the SM have been proposed to account for it. Explaining the origins
and properties of neutrino masses and DM continues to be an active
area of research.

After the discovery of the Higgs boson, which proved the existence of
a scalar field and provided important insights into the mechanism of
mass generation, the LHC has continued to search for additional scalar
particles. In fact, many Beyond the Standard Model (BSM) scenarios
include new scalar states. This is also the case of models addressing
the neutrino and DM problems, which typically require extended scalar
sectors. If these new states have masses and couplings within the
reach of the LHC, their signals may be hidden in the currently
existing searches or show up in the near future, and may appear at
lower or higher masses than the scalar found at the LHC at a mass of
125 GeV. If lighter states exist, they may be produced given the high
energies available at colliders. In fact, different experiments have
performed searches for low mass scalars in different
channels~\cite{OPAL:2002ifx,
  LEPWorkingGroupforHiggsbosonsearches:2003ing, ALEPH:2006tnd,
  CDF:2012wzv, CMS:2015ocq, CMS:2018cyk, CMS:2023yay, ATLAS:2018xad,
  CMS:2018rmh, CMS:2022goy, ATLAS:2022abz}.

The diphoton channel plays a crucial role in the search for new scalar
particles at the LHC. This final state allows for precise measurements
and clean experimental signatures, making it easier to isolate
potential signals of new scalar particles amidst background
noise. Interestingly, the CMS collaboration has been consistently
finding an excess over the SM prediction in this channel at a diphoton
invariant mass of $\sim 95$ GeV~\cite{CMS:2015ocq,CMS:2018cyk}. The
statistical support for this excess has been reinforced by recent
results obtained after the analysis of the full Run 2
dataset~\cite{CMS:2023yay}. The excess is maximal for a mass of $95.4$
GeV and has a local (global) significance of $2.9 \, \sigma$ ($1.3 \,
\sigma$) and can be interpreted as the production via gluon fusion and
subsequent decay of a new scalar state, $h_{95}$. It can be
parametrized numerically in terms of the $\mu_{\gamma\gamma}$ signal
strength, which normalizes the cross section of the process to the
analogous cross section for a Higgs-like state $H$ at the same mass. The
latest CMS result points to~\cite{CMS:2023yay, CMS:Moriond}
\begin{equation} \label{eq:CMS}
\mu^{\rm CMS}_{\gamma\gamma} = \frac{\sigma^{\rm CMS}(gg\to h_{95} \to\gamma\gamma)}{\sigma^{\rm SM}(gg\to H\to\gamma\gamma)} = 0.33^{+0.19}_{-0.12} \, .
\end{equation}

The ATLAS collaboration has also performed searches in the diphoton
channel, although with a lower sensitivity. A very mild excess in a
mass region compatible with that hinted by CMS was found in their Run
1 analysis~\cite{ATLAS:2018xad}. Their update including 140 fb$^{-1}$
of Run 2 data appeared recently~\cite{ATLAS:seminar}. Intriguingly,
the statistical significance of the excess increases in the new ATLAS
results. This can be attributed to the addition of more statistics as
well as to several improvements in the analysis. In particular, the
model-dependent analysis presented in this update hints at an excess,
curiously at $95.4$ GeV too, with a local significance of $1.7 \,
\sigma$. We note that this result is compatible with that of CMS, which
ATLAS cannot exclude.

The 95 GeV region is particularly interesting due to the existence of
other excesses hinting at similar mass values. LEP has reported an
excess in $b\bar{b}$ production at about 95 GeV with a local
significance of $2 \,
\sigma$~\cite{LEPWorkingGroupforHiggsbosonsearches:2003ing}. This
excess can be interpreted in terms of a new scalar state contributing
to the process $e^+e^- \to Z \, h_{95} \to Z \, b\bar{b}$, with a
signal strength given by~\cite{Azatov:2012bz,Cao:2016uwt}
\begin{equation}
\mu^{\rm LEP}_{bb}=\frac{\sigma^{\rm LEP}(e^+e^-\to Z \, h_{95} \to Z \, b\bar{b})}{\sigma^{\rm SM}(e^+e^-\to Z \, H\to Z \, b\bar{b})}=0.117 \pm 0.057 \, .
\end{equation}
 
Other searches for light scalars in CMS also gave a small excess in
the ditau channel $\mu_{\tau\tau}^{\rm CMS}=1.2\pm
0.5$~\cite{CMS:2022goy}. ATLAS has not published any ditau search in
this mass region, but has only provided results for scalar masses
above $200$ GeV~\cite{ATLAS:2017eiz}.

The 95 GeV excesses have received some attention
recently~\cite{Cao:2016uwt,
  Fox:2017uwr,Richard:2017kot,Haisch:2017gql,Biekotter:2017xmf,Domingo:2018uim,Liu:2018xsw,Cline:2019okt,Biekotter:2019kde,Aguilar-Saavedra:2020wrj,Heinemeyer:2018wzl,Heinemeyer:2018jcd,Biekotter:2021ovi,Biekotter:2021qbc,Heinemeyer:2021msz,Biekotter:2022jyr,Biekotter:2022abc,Biekotter:2023jld,Bonilla:2023wok,Azevedo:2023zkg}. We
interpret them in a variant of the Scotogenic model~\cite{Ma:2006km},
a well-motivated and economical BSM scenario that incorporates a
mechanism for the generation of neutrino masses and provides a
testable DM candidate. We thus consider the possibility that these
excesses are the first collider hints of a new BSM sector
addressing some of the most important open questions in particle
physics.

The rest of the manuscript is organized as
follows. Section~\ref{sec:model} introduces our model, whereas
Section~\ref{sec:inter} interprets the 95 GeV excesses in terms of a
new scalar state in the particle spectrum. The most relevant
experimental constraints are discussed in
Section~\ref{sec:constraints} and our numerical results, which prove
that our setup can accommodate the excesses, are presented in
Section~\ref{sec:num}. Other aspects of our scenario, such as neutrino
masses, DM and additional collider signatures, are discussed in
Section~\ref{sec:discussion}. Finally, we summarize our work in
Section~\ref{sec:summary}. An Appendix is also included with some
technical details.

\section{The Model}
\label{sec:model}

We consider a variant of the Scotogenic model~\cite{Ma:2006km} that
extends the SM particle content with $n_N$ generations of singlet
fermions $N_n$, $n = 1, \dots, n_N$, and $n_\eta$ doublet scalars
$\eta_a$, $a = 1, \dots, n_\eta$. These fields are assumed to be odd
under a new \z2 symmetry, under which all the SM states are even. This
generalizes the original Scotogenic model~\cite{Escribano:2020iqq},
which corresponds to $n_N=3$ and $n_\eta=1$. The $\eta_a$ doublets can
be decomposed in terms of their $\rm SU(2)_L$ components as
\begin{equation}
  \eta_a = \left( \begin{array}{c}
    \eta_a^+ \\
    \eta_a^0 \end{array} \right) \, .
\end{equation}
In addition, we include a real singlet scalar $S$. The lepton and
scalar particle content of the model is summarized in
Tab.~\ref{tab:ParticleContent}.

\begin{table}[h!]
  \centering
  \begin{tabular}{|c|c||ccc|c|}
    \hline
    \textbf{Field} & \textbf{Generations} & $\boldsymbol{\rm SU(3)_c}$ & $\boldsymbol{\rm SU(2)_L}$ & $\boldsymbol{\rm U(1)_Y}$ & $\boldsymbol{\mathbb{Z}_2}$ \\
    \hline
    $\ell_L$       & $3$                    & 1                    & 2                    & -1/2                & +                       \\
    $e_R$          & $3$                    & 1                    & 1                    & -1                  & +                       \\
    $N$            & $n_N$                  & 1                    & 1                    & 0                   & -                       \\
    \hline
    $H$            & $1$                    & 1                    & 2                    & 1/2                 & +                       \\
    $\eta$         & $n_\eta$                & 1                    & 2                    & 1/2                 & -                       \\
    $S$            & $1$                    & 1                    & 1                    & 0                   & +                       \\
    \hline
  \end{tabular}
  \caption{Lepton and scalar particle content of the model and their representations under gauge and global symmetries.
  \label{tab:ParticleContent}}
\end{table}

The Yukawa Lagrangian of the model includes the terms
\begin{eqnarray} \label{eq:Lyuk}
  \mathcal{L} \supset y_{n a \alpha} \, \overline{N}_n \widetilde{\eta}_a^\dagger \ell_L^\alpha - \kappa_{nm} \, S \, \overline{N^c}_n N_m - \frac{1}{2} \left(M_N\right)_{nn} \overline{N^c}_n N_n + \hc \, ,
\end{eqnarray}
where $n,m=1,\dots,n_N$, $a=1,\dots,n_\eta$ and $\alpha=1,2,3$ are
generation indices. The Yukawa coupling $y$ is an $n_N \times n_\eta
\times 3$ object while $\kappa$ and $M_N$ are $n_N \times n_N$
symmetric matrices. $M_N$ has been chosen diagonal without loss of
generality. Finally, we define $\widetilde{\eta} = i \sigma_2
\eta^*$. The scalar potential of the model can be written as
\begin{equation} \label{eq:V}
  \mathcal V = \mathcal V_H + \mathcal V_\eta + \mathcal V_S + \mathcal V_{\rm mix} \, ,
\end{equation}
with
\begin{align}
  \mathcal V_H =& m_H^2 \, H^\dagger H + \frac{1}{2} \lambda_1 \, \left(H^\dagger H\right)^2 \, , \\
  \mathcal V_\eta =& \left(m_\eta^2\right)_{a a} \eta_a^\dagger \eta_a + \frac{1}{2} \lambda_2^{a b c d} \left(\eta_a^\dagger \eta_b\right)\left(\eta_c^\dagger \eta_d \right) \, , \\
  \mathcal V_S =& \frac{1}{2} m_S^2 \, S^2 + \frac{1}{3} \mu_S \, S^3 + \frac{1}{4} \lambda_S \, S^4 \, , \\
  \mathcal V_{\rm mix} =& \lambda_3^{a b} \left(H^\dagger H\right)\left(\eta_a^\dagger \eta_b\right)+\lambda_4^{a b}\left(H^\dagger \eta_a\right)\left(\eta_b^\dagger H\right) +\frac{1}{2} \left[\lambda_5^{a b} \left(H^\dagger \eta_a\right)\left(H^\dagger \eta_b\right)+ \hc \right] \nonumber \\
  & + \mu_H \, H^\dagger H S + \frac{1}{2} \lambda_{HS} \, H^\dagger H S^2 + \mu_\eta^{a b} \, \eta_a^\dagger \eta_b \, S + \frac{1}{2} \lambda_{\eta S}^{a b} \, \eta_a^\dagger \eta_b \, S^2 \, .
\end{align}
Here all the indices are $\eta$ generation indices and then
$m_\eta^2$, $\lambda_{3,4,5}$, $\mu_\eta$ and $\lambda_{\eta S}$ are
$n_\eta \times n_\eta$ matrices, while $\lambda_2$ is an $n_\eta
\times n_\eta \times n_\eta \times n_\eta$ object. We also note that
$\lambda_5$ must be symmetric whereas $\lambda_{3,4}$, $\mu_\eta$ and
$\lambda_{\eta S}$ are Hermitian. Again, $m_\eta^2$ will be assumed to
be diagonal without any loss of generality.

\subsection{Symmetry breaking, scalar masses and mixings}

We will assume that the vacuum of our model is given by
\begin{equation} \label{eq:vevs}
\langle H^0 \rangle = \frac{v}{\sqrt{2}} \, , \quad \langle \eta^0_a \rangle = 0 \, , \quad \langle S \rangle = v_S \, .
\end{equation}
The vacuum expectation value (VEV) $v$ breaks the electroweak symmetry
in the usual way. In contrast, the \z2 Scotogenic parity remains
exactly conserved due to $\langle \eta_a^0 \rangle = 0$. The VEV
configuration in Eq.~\eqref{eq:vevs} imposes some conditions on the
scalar potential parameters due to the minimization equations
\begin{align}
\left.\frac{\partial V}{\partial H}\right|_{\langle H \rangle = \frac{v}{\sqrt{2}} \, , \, \langle S \rangle = v_S} =& m_H^2 v + \mu_H v \, v_S + \frac{1}{2} \lambda_1 v^3 + \frac{1}{2} \lambda_{HS} v_S^2 v =0 \, , \label{eq:tad1}  \\
\left.\frac{\partial V}{\partial S}\right|_{\langle H \rangle = \frac{v}{\sqrt{2}} \, , \, \langle S \rangle = v_S} =& m_S^2 v_S + \mu_S v_S^2 + \frac{1}{2} \mu_H v^2 + \lambda_S v_S^3 + \frac{1}{2}\lambda_{HS} v^2 \, v_S = 0 \, . \label{eq:tad2}
\end{align}
After symmetry breaking, the real component of the neutral $H^0$ field
mixes with the real $S$ field. In the basis $\mathcal H = \left\{ S ,
\text{Re}(H^0) \right\}$ their mass matrix reads
\begin{eqnarray}
\mathcal{M}_{\mathcal H}^2 = \left(
\begin{matrix}
m_S^2 + 2 \mu_S v_S + \frac{1}{2}\lambda_{HS} \, v^2 + 3 \lambda_S v_S^2 & \mu_H v + \lambda_{HS} \, v \, v_S \\
\mu_H v + \lambda_{HS} \, v \, v_S  & m_H^2 + \mu_H v_S + \frac{3}{2} \lambda_1 v^2 + \frac{1}{2}\lambda_{HS} \, v_S^2
\end{matrix} 
\right) \, .
\end{eqnarray}
After application of Eqs.~\eqref{eq:tad1} and \eqref{eq:tad2}, solved
for $m_H^2$ and $m_S^2$, this matrix becomes
\begin{eqnarray}
\mathcal{M}_{\mathcal H}^2 = \left(
\begin{matrix}
v_S \left( \mu_S + 2 \lambda_S v_S \right) - \frac{\mu_H v^2}{2 v_S} &  \mu_H v + \lambda_{HS} \, v \, v_S \\
\mu_H v + \lambda_{HS} \, v \, v_S & \lambda_1 v^2
\end{matrix} 
\right) \, .
\end{eqnarray}
It can be brought to diagonal form as $V_{\mathcal H} \,
\mathcal{M}^2_{\mathcal H} \, V_{\mathcal H}^T =
\widehat{\mathcal{M}}^2_{\mathcal H} = \text{diag} \left(m^2_{h_1} ,
m^2_{h_2}\right)$, where $h_1$ and $h_2$ are mass eigenstates and
\begin{equation}
  V_{\mathcal H} = \left( \begin{matrix}
    \cos \alpha & \sin \alpha \\
    - \sin \alpha & \cos \alpha \end{matrix} \right) \, ,
\end{equation}
with
\begin{equation}
\tan 2\alpha = \frac{2 \, \left( \mathcal{M}^2_{\mathcal H} \right)_{12}}{\left( \mathcal{M}^2_{\mathcal H} \right)_{11}-\left( \mathcal{M}^2_{\mathcal H} \right)_{22}} \, .
\end{equation}
This mixing angle between the singlet and doublet scalars plays a
central role in the phenomenology of our model, as we will explain in the following Subsection. We focus now on the \z2-odd scalars $\eta^+_a$ and $\eta^0_a$. We decompose the neutral components of the $\eta_a$ doublets as
\begin{equation}
  \eta_a^0 = \frac{1}{\sqrt{2}} (\eta_{R_a} + i \eta_{I_a}) \, ,
\end{equation}
and they do not mix if we assume that CP is conserved in the scalar sector. This can be easily achieved if all the parameters in the scalar potential are real. Again, after electroweak symmetry breaking, the $n_\eta \times n_\eta$ mass matrices are given by
\begin{align}
  \bigl( \mathcal{M}_{\eta_R}^2 \bigr)_{ab} & = \left(m_\eta^2\right)_{ab} + \left( \lambda_3^{ab} + \lambda_4^{ab} + \lambda_5^{ab} \right) \frac{v^2}{2} + \frac{v_S^2}{2} \lambda_{\eta S}^{ab} + \mu_\eta^{ab} v_S \, , \\
  \bigl( \mathcal{M}_{\eta_I}^2 \bigr)_{ab} & = \left(m_\eta^2\right)_{ab} + \left( \lambda_3^{ab} + \lambda_4^{ab} - \lambda_5^{ab} \right) \frac{v^2}{2} + \frac{v_S^2}{2} \lambda_{\eta S}^{ab} + \mu_\eta^{ab} v_S \, , \\
  \bigl( \mathcal{M}_{\eta^+}^2 \bigr)_{ab} & = \left(m_\eta^2\right)_{ab} + \lambda_3^{ab}\frac{v^2}{2} + \frac{v_S^2}{2} \lambda_{\eta S}^{ab} + \mu_\eta^{ab} v_S \, . 
  \label{eq:etamass}
\end{align}
Notice that the mass matrices for the real and imaginary components are the same in the limit in which all the elements of $\lambda_5$ vanish.

\subsection{Scalar couplings}

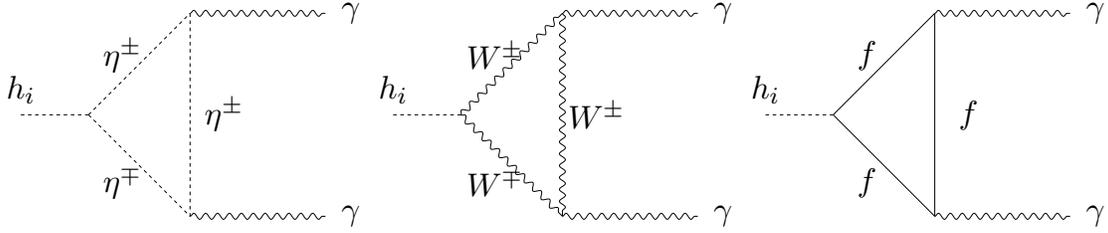
\begin{figure}
  \begin{center}
    \scalebox{0.45}{
      \begin{tikzpicture}
	\begin{scope}[thick] 
	  \draw[thick, dashed] (3,0)--(5,0);
	  \draw[thick, dashed] (5,0)--(8,3);
	  \draw[thick, dashed] (5,0)--(8,-3);
	  \draw[thick, dashed] (8,3)--(8,-3);
	  \draw[thick, photon] (8,3)--(12,3);
	  \draw[thick, photon] (8,-3)--(12,-3);
	  
	  \node[black,scale=2.5] at (3.0,0.75) {{$h_i$}};
	  \node[black,scale=2.5] at (6.0,1.75) {{$\eta^{\pm}$}};
	  \node[black,scale=2.5] at (6.0,-2.00) {{$\eta^{\mp}$}};
	  \node[black,scale=2.5] at (9,0.0) {{$\eta^{\pm}$}};
	  \node[black,scale=2.5] at (12.75,3) {{$\gamma$}};
	  \node[black,scale=2.5] at (12.75,-3) {{$\gamma$}};
	  %	\node[fill,circle,scale=1.0] at (2,0) {};%
	  %	\node[black,scale=2.0] at (-1.0,-3) {{$g_{h_i\eta\eta}$}};
	  %	\draw[->]        (1.5,-0.5)   -- (-0.75,-2.5);
	  
	  \draw[thick, dashed] (14,0)--(16,0);
	  \draw[thick, photon] (16,0)--(19,3);
	  \draw[thick, photon] (16,0)--(19,-3);
	  \draw[thick, photon] (19,3)--(19,-3);
	  \draw[thick, photon] (19,3)--(23,3);
	  \draw[thick, photon] (19,-3)--(23,-3);	
	  \node[black,scale=2.5] at (14.0,0.75) {{$h_i$}};
	  \node[black,scale=2.5] at (17.0,1.75) {{$W^{\pm}$}};
	  \node[black,scale=2.5] at (17.0,-2.00) {{$W^{\mp}$}};
	  \node[black,scale=2.5] at (20,0.0) {{$W^{\pm}$}};
	  \node[black,scale=2.5] at (23.75,3) {{$\gamma$}};
	  \node[black,scale=2.5] at (23.75,-3) {{$\gamma$}};		
	  
	  \draw[thick, dashed] (25,0)--(27,0);
	  \draw[thick, -] (27,0)--(30,3);
	  \draw[thick, -] (27,0)--(30,-3);
	  \draw[thick, -] (30,3)--(30,-3);
	  \draw[thick, photon] (30,3)--(34,3);
	  \draw[thick, photon] (30,-3)--(34,-3);
	  \node[black,scale=2.5] at (25.0,0.75) {{$h_i$}};
	  \node[black,scale=2.5] at (28.0,1.75) {{$f$}};
	  \node[black,scale=2.5] at (28.0,-2.00) {{$f$}};
	  \node[black,scale=2.5] at (31,0.0) {{$f$}};
	  \node[black,scale=2.5] at (34.75,3) {{$\gamma$}};
	  \node[black,scale=2.5] at (34.75,-3) {{$\gamma$}};		
	  
	\end{scope}
      \end{tikzpicture}
    }
  \end{center}
  \caption{1-loop contributions to $h_i \to \gamma\gamma$.
    \label{fig:h_gammagamma}}
\end{figure}

The couplings of the $h_1$ and $h_2$ scalars to the SM fermions are
determined by the $\alpha$ mixing angle which, as discussed below,
will be constrained to be small. We note that the singlet $S$ does not
have a Yukawa term with the SM fermions. Hence, it can only couple to
them via mixing with the SM Higgs $H$. As a result of this, one of the
mass eigenstates ($h_1$, the mostly-singlet one) couples to the SM
fermions proportional to $\sin\alpha$ while the other mass eigenstate
($h_2$, the SM-like one) couples proportional to $\cos\alpha$. The
same happens for the coupling to $WW$ and $ZZ$ bosons and the loop
coupling to gluons. However, the 1-loop couplings to $\gamma\gamma$
and $\gamma Z$ will be affected by the new particle content present in
our model, in particular the $\eta$ doublets. In
Fig.~\ref{fig:h_gammagamma} we can see the different contributions to
the loop decays $h_i \to \gamma\gamma$. Similar diagrams can be drawn
for the $\gamma Z$ final state. On the one hand the Higgses will couple
to $\gamma\gamma/Z$ through the SM loops with $W$ bosons and fermions,
with the largest contribution among the fermions given by the top
quark. On the other hand, the charged $\eta$ states can also run into
the loop and contribute to the decay through the $g_{h_i\eta\eta}$
coupling, given by
\begin{equation} \label{eq:ghetaeta}
g_{h_i\eta\eta} = \left(V_{\mathcal H}\right)_{i1} (\lambda_{\eta S} \, v_S + \mu_\eta) + \left(V_{\mathcal H}\right)_{i2} \, \lambda_3 \, v \, ,
\end{equation}
or, equivalently,
\begin{align}
g_{h_1\eta\eta} =& \cos\alpha \left(\lambda_{\eta S} \, v_S + \mu_\eta \right) + \sin\alpha \, \lambda_3 \, v \, , \label{eq:ghetaeta1} \\
g_{h_2\eta\eta} =& - \sin\alpha \left(\lambda_{\eta S} \, v_S + \mu_\eta \right) + \cos\alpha \, \lambda_3 \, v \, . \label{eq:ghetaeta2}
\end{align}

\subsection{Neutrino masses}
\label{subsec:numass}

After symmetry breaking, the $n_N \times n_N$ mass matrix of the
singlet fermions is given by
\begin{equation}
  \mathcal M_N = M_N + 2 \, \kappa \, v_S \, .
\end{equation}
As already explained, one can take the matrix $M_N$ to be diagonal
without loss of generality. We will further assume that $\kappa$ is
diagonal too. Then, the singlet fermion masses are simply given by
$m_{N_n} = \left(\mathcal M_N\right)_{nn} = \left(M_N\right)_{nn} + 2
\, \kappa_{nn} \, v_S$.

\begin{figure}
  \centering
  \includegraphics[width=.5\textwidth]{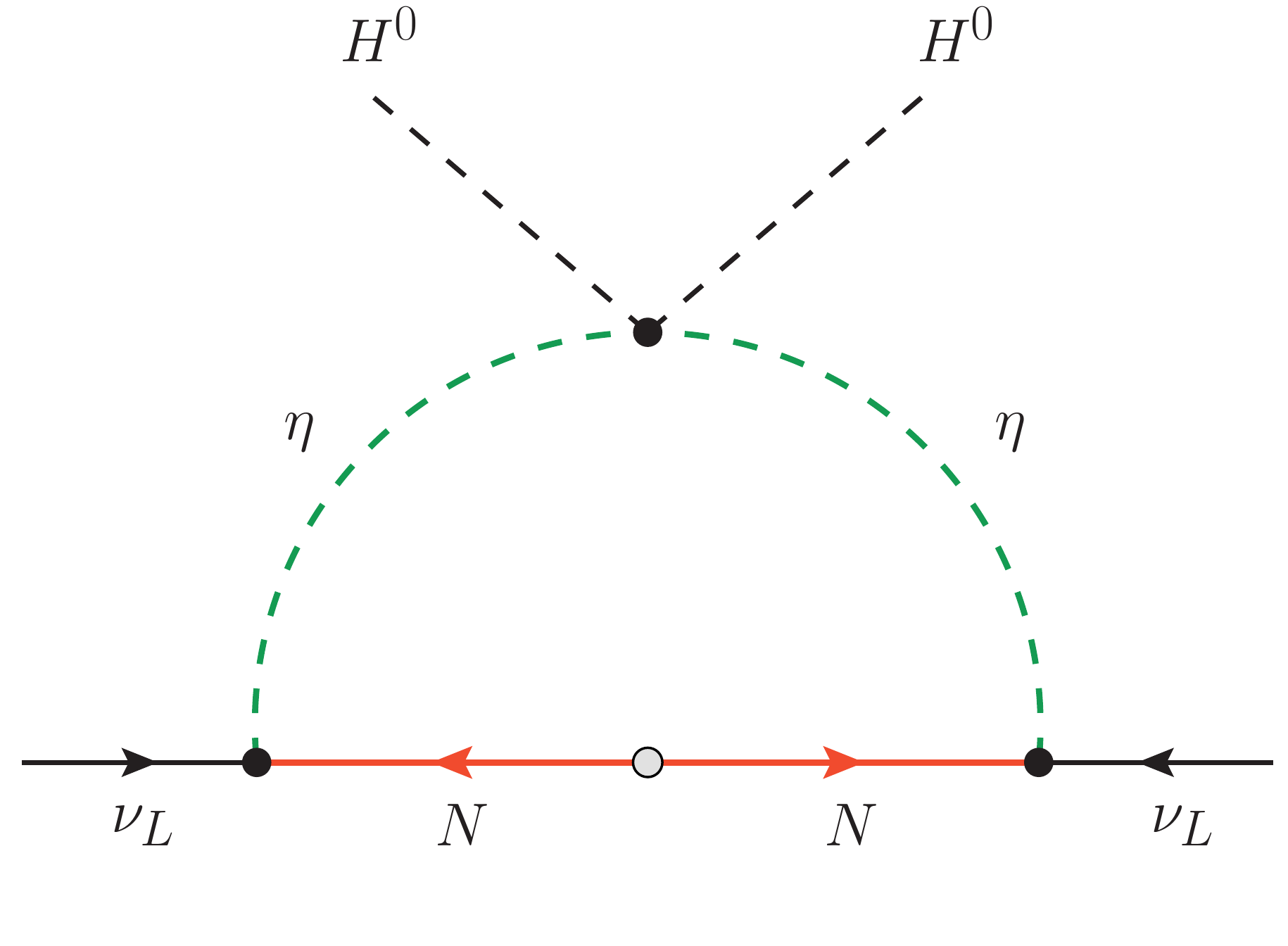}
  \caption{Neutrino mass generation in our model.
  \label{fig:numass}}
\end{figure}

The simultaneous presence of the $y$ and $\lambda_5$ couplings and the
$M_N$ Majorana mass term (or the $\kappa$ coupling) leads to explicit
lepton number violation. Neutrino masses vanish at tree-level due to
the \z2 symmetry of the model, that forbids a neutrino Yukawa
interaction with the SM Higgs doublet. However, neutrinos acquire
non-zero Majorana masses at the 1-loop, as shown in
Fig.~\ref{fig:numass}. This mechanism is exactly the same as in the
original Scotogenic model~\cite{Ma:2006km}, although our scenario
includes a variable number of $N$ and $\eta$ fields. The general
expression for the light neutrinos Majorana mass matrix can be found
in~\cite{Escribano:2020iqq} where it is particularized for specific
$(n_N,n_\eta)$ cases.

\section{Interpretation of the 95 GeV excesses}
\label{sec:inter}

In the following we will assume that the lightest \z2-even scalar in
our model, $h_1$, has a mass of 95 GeV and is thus identified with the
scalar resonance hinted by CMS, ATLAS and LEP precisely this
energy scale. Therefore, $h_2$ is identified with the 125 GeV Higgs
discovered at the LHC. In summary,
\begin{equation}
  h_1 \equiv h_{95} \, , \quad h_2 \equiv h_{125} \, .
\end{equation}
We should then study whether our model can accommodate the
experimental hints at 95 GeV. In other words, we must determine the
regions in the parameter space leading to a $h_1$ diphoton signal
strength in agreement with Eq.~\eqref{eq:CMS} that also comply with
the existing experimental constraints. We will also explore the
possibility to simultaneously explain the other anomalies at 95 GeV,
in the $b \bar{b}$ and ditau channels.

The signal strength for the diphoton channel is given in our model by
\begin{equation}
\mu_{\gamma\gamma} = \frac{\sigma(gg\to h_1)}{\sigma_{\rm SM}(gg\to H)} \times \frac{{\rm BR}(h_1\to\gamma\gamma)}{{\rm BR}_{\rm SM}(H\to\gamma\gamma)} = \sin^2\alpha \, \frac{{\rm BR}(h_1\to\gamma\gamma)}{{\rm BR}_{\rm SM}(H\to\gamma\gamma)} \, ,
\end{equation}
where we normalize again to the SM values, the usual suppression by
the $\alpha$ mixing angle has been taken into account and
${\rm BR}(h_1\to \gamma\gamma)$ is the $h_1\to \gamma\gamma$ branching
ratio in our model. This is modified with respect to the predicted
value for a Higgs-like state with a mass of 95 GeV due to the presence
of the $\eta$ doublets. The decay width of a CP-even scalar to two
photons has been studied in great
detail~\cite{Djouadi:2005gi,Djouadi:2005gj,Staub:2016dxq}. With
$n_\eta$ generations of $\eta$ doublets and assuming diagonal
$g_{h_i\eta\eta}$ couplings, it is given by
\begin{equation}
\Gamma(h_i\to \gamma\gamma) = \frac{G_F\alpha^2 m_{h_i}^3}{128\sqrt{2}\pi^3}\left| \sum_f N_c Q_f^2 g_{h_iff} \, A_{1/2}(\tau_f) + g_{h_iWW} \, A_1(\tau_W) +\sum_a\frac{v}{2m^2_{\eta_a}}g_{h_i\eta\eta}^{aa} \, A_0(\tau_\eta)\right|^2 \, ,
\end{equation}
where $\tau_k=m^2_{h_i}/4m^2_k$ and the $A_i(\tau)$ functions are
defined in Appendix~\ref{app:loop}. As we can see, the presence of the
$\eta$ doublets not only modifies the diphoton decay of the
mostly-singlet state, but it also affects to the diphoton decay of the
SM-like Higgs. As discussed below in Sec.~\ref{sec:constraints}, this
feature constrains the parameter space from the existing measurements
of the 125 GeV Higgs.

For the $b \bar{b}$ excess in LEP we must consider the signal strength
\begin{equation} \label{eq:mubb}
\mu_{bb}=\frac{\sigma(e^+e^-\to Z \, h_1)}{\sigma_{\rm SM}(e^+e^-\to Z \, H)} \times \frac{{\rm BR}(h_1\to b\bar{b})}{{\rm BR}_{\rm SM}(H\to b\bar{b})}=\sin^2\alpha \, \frac{\Gamma(h_1\to b\bar{b})/\Gamma^{\rm tot}}{\Gamma_{\rm SM}(H\to b\bar{b})/\Gamma^{\rm tot}_{\rm SM}} = \sin^4\alpha \, \frac{\Gamma^{\rm tot}_{\rm SM}}{\Gamma^{\rm tot}} \, .
\end{equation}
In this case, the mixing angle not only suppresses the production
cross section, but also the decay width of $h_1 \to b\bar{b}$, which
can only take place via singlet-doublet mixing.

As we can see from both signal strengths, the main features of this
model to explain the signals are, on the one hand, the reduced
couplings to fermions and vector bosons given by the mixing of the
singlet state with the doublet, that introduce powers of $\sin\alpha$
in the observables of interest. This allows (i) to evade the existing
limits from LEP and the LHC on light scalars with masses below 125 GeV
decaying into SM states and, (ii) to easily accomodate the correct
range of values to explain the $b\bar{b}$ excess at LEP. This occurs
easily because a singlet of a mass of 95 GeV with a small admixture
with the doublet will predominantly decay into a $b\bar{b}$ pair. On
the other hand, the singlet couples directly to the $\eta$
doublets. This induces, via loops, the decay into a pair of
photons. This feature allows to explain in a natural way the diphoton
rate at CMS. We can see that the features for both
excesses have different origins, the $b\bar{b}$ signal comes through
the mixing with the doublet state while the diphoton signal is mainly
driven by the singlet couplings. In order to accomodate both signals,
the interplay between the singlet and doublet components of $h_1$ must
be looked for.

Finally, the signal strength for $\tau^+\tau^-$ is exactly the same as
that for $b\bar{b}$ in Eq.~\eqref{eq:mubb}. In fact, our model
predicts $\mu_{bb}=\mu_{\tau\tau}$. This obviously precludes our model
from explaining the CMS ditau excess in the same region of parameter
space that explains the LEP $b\bar{b}$ excess. In fact, as we will see
below, the value of the mixing angle $\alpha$ required to explain the
ditau excess is too large, already excluded by Higgs data. Thefore,
the CMS ditau excess will be interpreted as an upper limit instead.

\section{Constraints}
\label{sec:constraints}

The presence of a 95 GeV scalar that could lead to an explanation to
the anomalies in the data is subject to different constraints.

First of all there are some theoretical constraints that affect the
parameters of our model. Such constraints involve mainly the
parameters of the scalar potential. First of all, we demand all the
quartic couplings in the potential to be below $\sqrt{4\pi}$ in order
to ensure perturbativity. Furthermore, we demand the potential to be
bounded from below to ensure that we have a stable global
minimum. This requirement is rather complicated in the presence of
many scalar fields so we apply a \textit{copositivity} requirement as
described in the Appendix of Ref.~\cite{Escribano:2020iqq}. Although
being overconstraining, once the potential passes this requirement, it
is guaranteed to be bounded from below.

Furthermore, several experimental searches are sensitive to the
spectrum of our singlet extension of the Scotogenic model. In that
sense, the most important searches are the ones provided by
colliders. Since the Higgs sector gets modified with respect to the
one of the SM, one must ensure that all our predictions are in
agreement with the existing collider measurements. We remind the
reader that we have adopted a setup characterized by a light scalar
around 95 GeV and a SM-like Higgs boson at 125 GeV. In order to take
into account the bounds on the 95 GeV scalar we make use of the public
code \texttt{HiggsBounds-v.6}~\cite{Bechtle:2008jh, Bechtle:2011sb,
  Bechtle:2012lvg, Bechtle:2013wla, Bechtle:2015pma, Bechtle:2020pkv,
  Bahl:2021yhk}, integrated now in the public code
\texttt{HiggsTools}~\cite{Bahl:2022igd}. This code compares the
potential signatures of such a scalar against BSM scalar searches
performed at the LHC. A point in the parameter space of our model
would be excluded if its signal rate for the most sensitive channel
excedes the observed experimental limit at the 95\% confidence level.

Moreover, as a 125 GeV SM-like Higgs boson has been observed at the
LHC, we ask the second scalar state to be in agreement with the
experimental measurements of its signal rates using the public code
\texttt{HiggsSignals-v.3}~\cite{Bechtle:2013xfa, Stal:2013hwa,
  Bechtle:2014ewa, Bechtle:2020uwn} that is also part of
\texttt{HiggsTools}~\cite{Bahl:2022igd}. This code constructs a
$\chi^2$ function using the different data from the measured cross
sections at the LHC involving the measured 125 GeV Higgs boson. For
that purpose we provide the code with the rescaled effective couplings
for the different sensitive channels in our model. Once the $\chi^2$
function is built for a point of the parameter space we compare it
with the result of the fit for a $125.09$ GeV SM-like Higgs boson with
\texttt{HiggsSignals-v.3}, $\chi^2_{\rm{SM},125}=152.49$, and impose
that the difference between the calculated $\chi^2_{125}$ and
$\chi^2_{\rm{SM},125}$ is less than $2\sigma$ away from the LHC
measurements in order to consider a point as experimentally
allowed. Since we perform a two-dimensional analysis of the parameter
space, our consideration for a point to be allowed becomes
$\Delta\chi^2_{125} = \chi^2_{125} - \chi^2_{\rm{SM},125} \leq 6.18$.

Another point to have into consideration is the fact that the presence
of an $\rm SU(2)_L$ scalar doublet can induce sizeable contributions
to the electroweak precision observables. In particular, the oblique
parameters $S$, $T$ and $U$ are generally affected by the presence of
these particles, but these strongly depend on the scalar
masses~\cite{Barbieri:2006dq, Abada:2018zra}. When the CP-even and
CP-odd \z2-odd neutral states are mostly degenerate, or equivalently
when the entries of the $\lambda_5$ matrix are small, the $T$
parameter imposes a restrictive bound over the difference in masses
between the charged and neutral states, $\Delta m
(\eta^+,\eta^0)=|m_{\eta^+}-m_{\eta^0}|\lesssim 140$
GeV~\cite{Abada:2018zra}. Charged particles are also heavily
constrained by different searches at colliders. However, these
searches assume specific decay modes. In our case the decay of the
charged $\eta^\pm$ scalar takes place via electroweak couplings as
$\eta^\pm \to \eta^0 W^\pm$. Since all decay chains must include the
DM state, this eventually leads to missing transverse energy and
leptons or jets in the final state. For that purpose we impose a
conservative limit on the charged particles given by the LEP
experiment of about $m_{\eta^\pm}\gtrsim 100$ GeV. We impose this
bound even if a detailed analysis could show that the limits might be
weaker in some specific configurations, due to the decay modes and
mass differences. Such a detailed analysis is out of the scope of this
paper, that just aims at showing that our model can accommodate the 95
GeV excesses. The LHC has also performed searches looking for charged
particles that decay into a neutral one and different
objects~\cite{ATLAS:2022hbt, CMS:2023qhl}. Although the current limits
on charged particles can reach high values of the mass, they are again
strongly dependent on the mass splitting between the charged and
neutral states, making the searches almost not sensitive for
differences lower than $\Delta m (\eta^+,\eta^0)\lesssim 60$~
GeV. Furthermore, there are searches that look for charged particles
that decay into neutral states that are close in mass, producing soft
objects as final state~\cite{ATLAS:2019lng, CMS:2021edw}. These
searches aim to cover the gap in mass values of the previous analyses
for charged particles. Their sensitivity is maximized for mass
differences of order $\Delta m (\eta^+,\eta^0)\sim$ 10 GeV, decreasing
for increased values of $\Delta m (\eta^+,\eta^0)$, until it reaches
$\Delta m (\eta^+,\eta^0)\sim$ 60 GeV, where the searches from
Refs.~\cite{ATLAS:2022hbt, CMS:2023qhl} are sensitive. For that
reason, we take the masses of the $\eta$ doublet in such a way that
fulfil both $S$, $T$ and $U$ parameters and the collider
constraints. This requirement can be achieved naturally in this model
according to Eq~\eqref{eq:etamass} since $\Delta m (\eta^+,\eta^0)$ is
driven by the couplings $\lambda^{ab}_4$ and $\lambda^{ab}_5$. The
first of these matrices has entries tipically smaller than 1, while
the second one is usually very small due to its link with neutrino
masses.

\section{Numerical results}
\label{sec:num}

\begin{figure*}
  \centering
  \includegraphics[width=.43\textwidth]{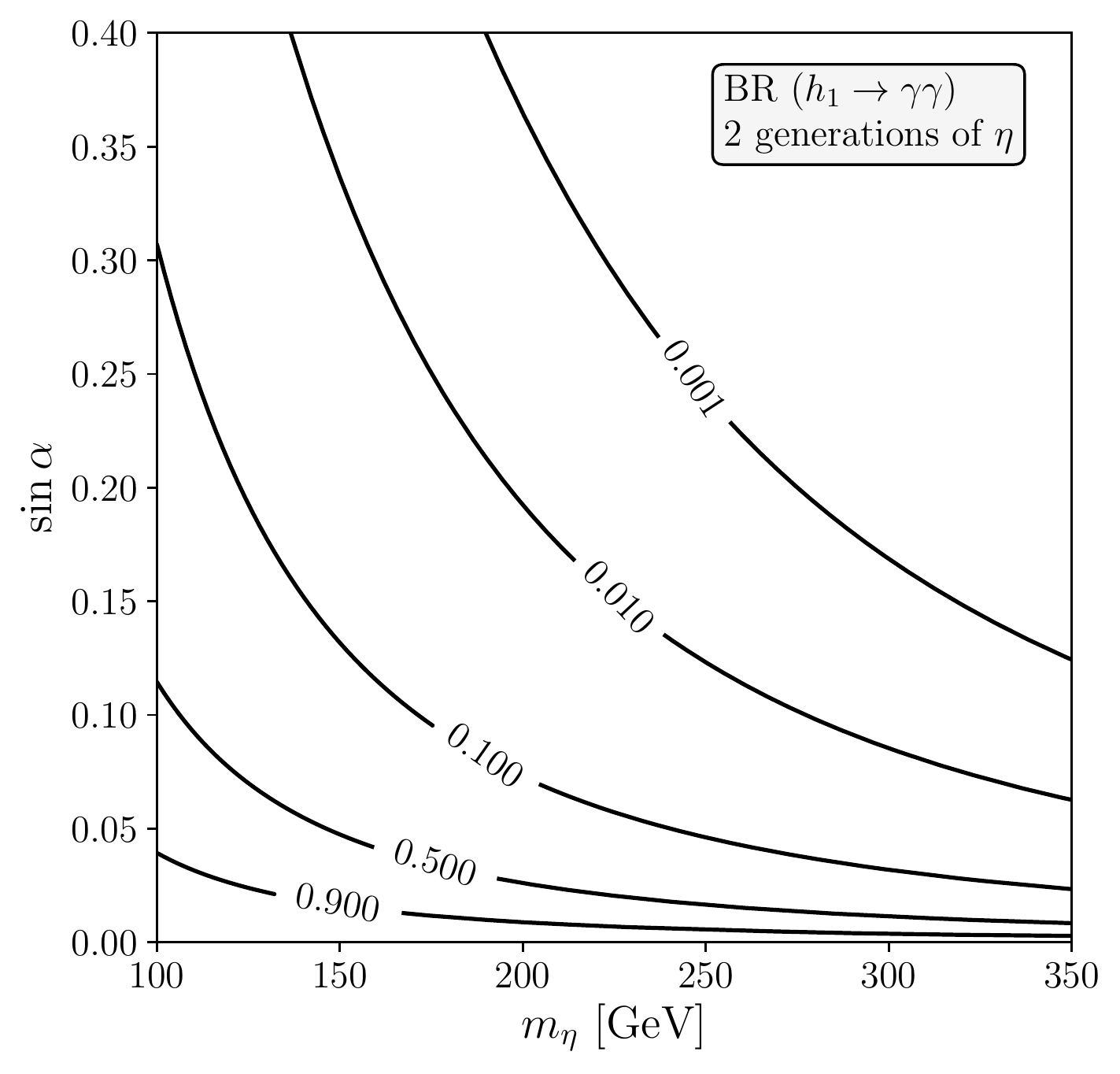}
  \includegraphics[width=.43\textwidth]{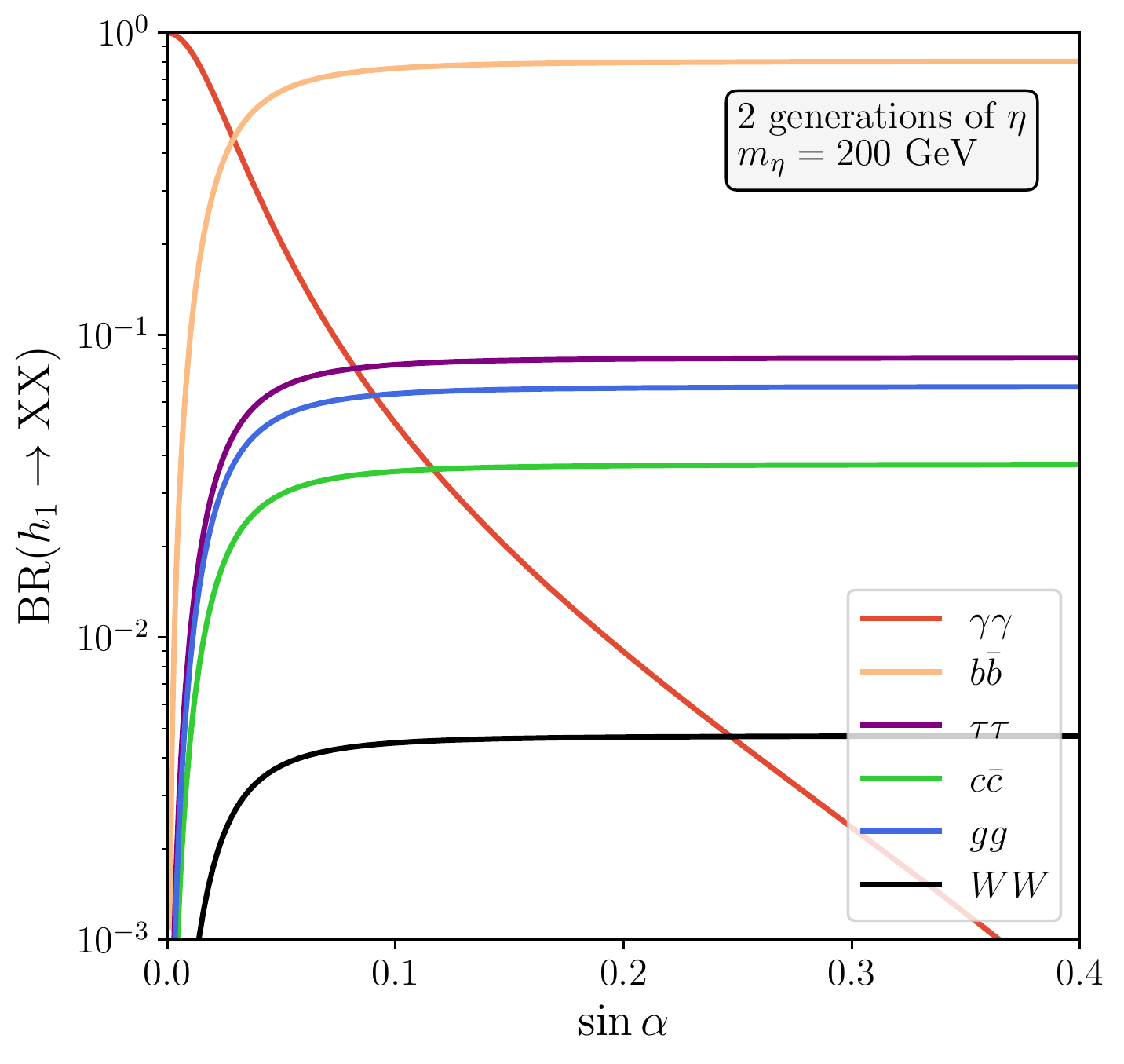}
  \caption{Branching ratios of the $h_1$ scalar into several
    final states for $n_\eta = 2$, $\overline{\lambda}_3 = 0.55$,
    $\overline{\lambda}_{\eta S} \, v_S = 500$ GeV and $\overline{\mu}_\eta = 1500$
    GeV. Left: contours of BR($h_1 \to
    \gamma \gamma$) in the $m_\eta - \sin \alpha$ plane. Right:
    BR($h_1 \to$ XX) as a function of $\sin \alpha$ for a fixed
    $m_\eta = 200$ GeV.
    \label{fig:h1BRs}}
\end{figure*}

We now show our numerical results. For the sake of simplicity, we will assume in the following that $\lambda_3$, $\mu_\eta$ and $\lambda_{\eta S}$ are proportional to the $n_\eta \times n_\eta$ identity matrix $\mathcal{I}_{n_\eta}$, that is, $X = \overline{X} \, \mathcal{I}_{n_\eta}$, with $X = \lambda_3$, $\mu_\eta$, $\lambda_{\eta S}$. Fig.~\ref{fig:h1BRs} shows our results for the $h_1$ decay width and branching ratios
into several final states. This figure has been made with the specific
choice $n_\eta = 2$, fixing also $\overline{\lambda}_3 = 0.55$, $\overline{\lambda}_{\eta S}
\, v_S = 500$ GeV and $\overline{\mu}_\eta = 1500$ GeV. The left-hand side of this figure shows contours of
BR($h_1 \to \gamma \gamma$) in the $m_\eta - \sin \alpha$ plane. One
can see that BR($h_1 \to \gamma \gamma$) decreases with $m_\eta$, as
expected, and gets enhanced for low values of $\sin \alpha$. In fact,
the branching ratio into the diphoton final state can be of order $1$
for very low values of $\sin \alpha$. This behavior is also
illustrated on the right-hand panel of the figure, which shows the
dependence of the different BR($h_1 \to$ XX) on $\sin \alpha$ for a
fixed $m_\eta = 200$ GeV. The enhancement is caused by the strong
suppression of all the other channels, which have negligible branching
ratios for low values of $\sin \alpha$. This is simply due to the fact
that the $h_1 \approx S$ decay into SM states can only take place via
singlet-doublet mixing. It is important to notice that for values of $\sin\alpha\gtrsim 0.05$ the branching ratio to a $b\bar{b}$ pair becomes predominant favouring the LEP signal as was explained in Sec.~\ref{sec:inter}.

\begin{figure*}
  \centering
  \includegraphics[width=.43\textwidth]{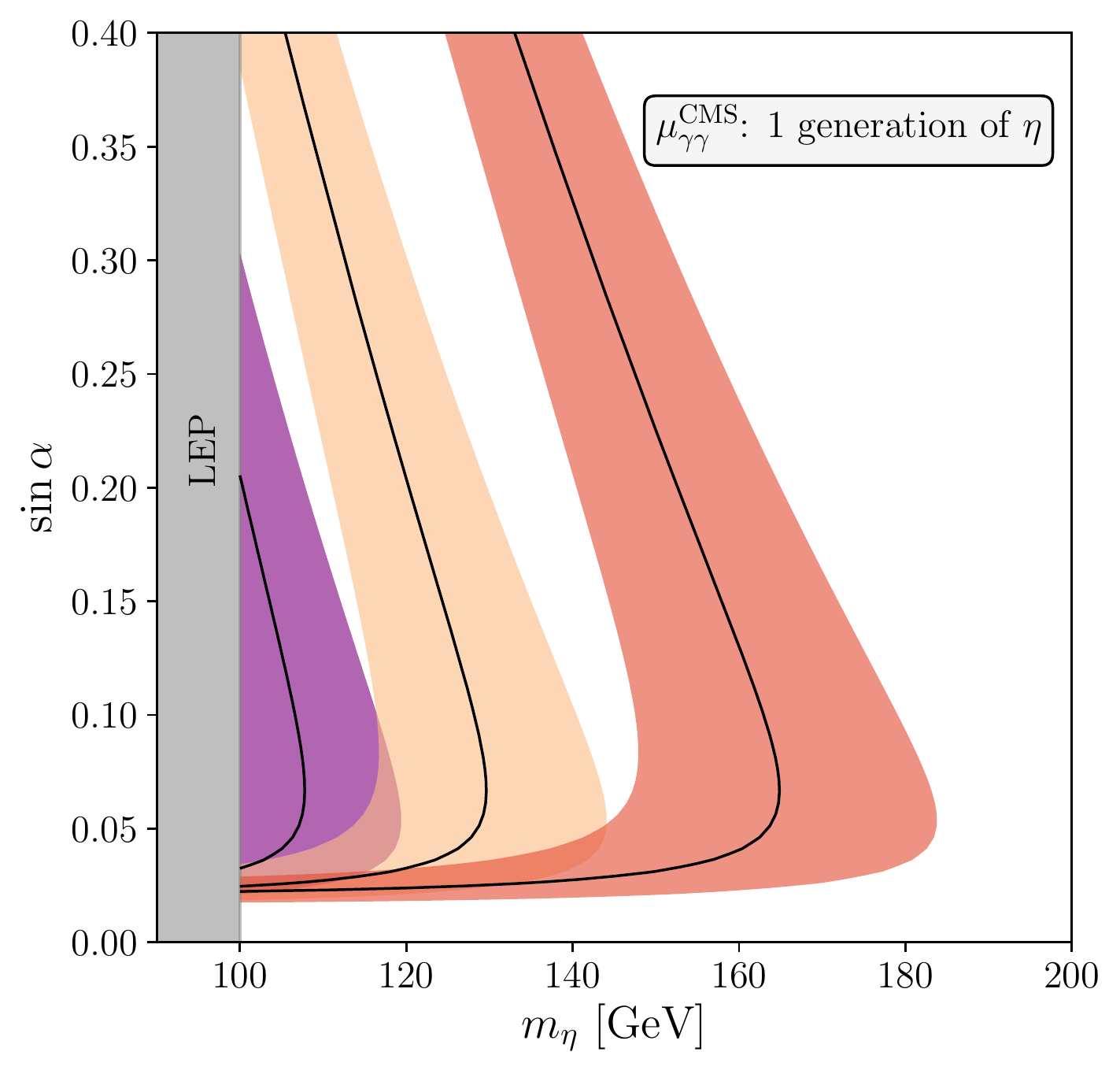}\hspace{1cm}
  \includegraphics[width=.43\textwidth]{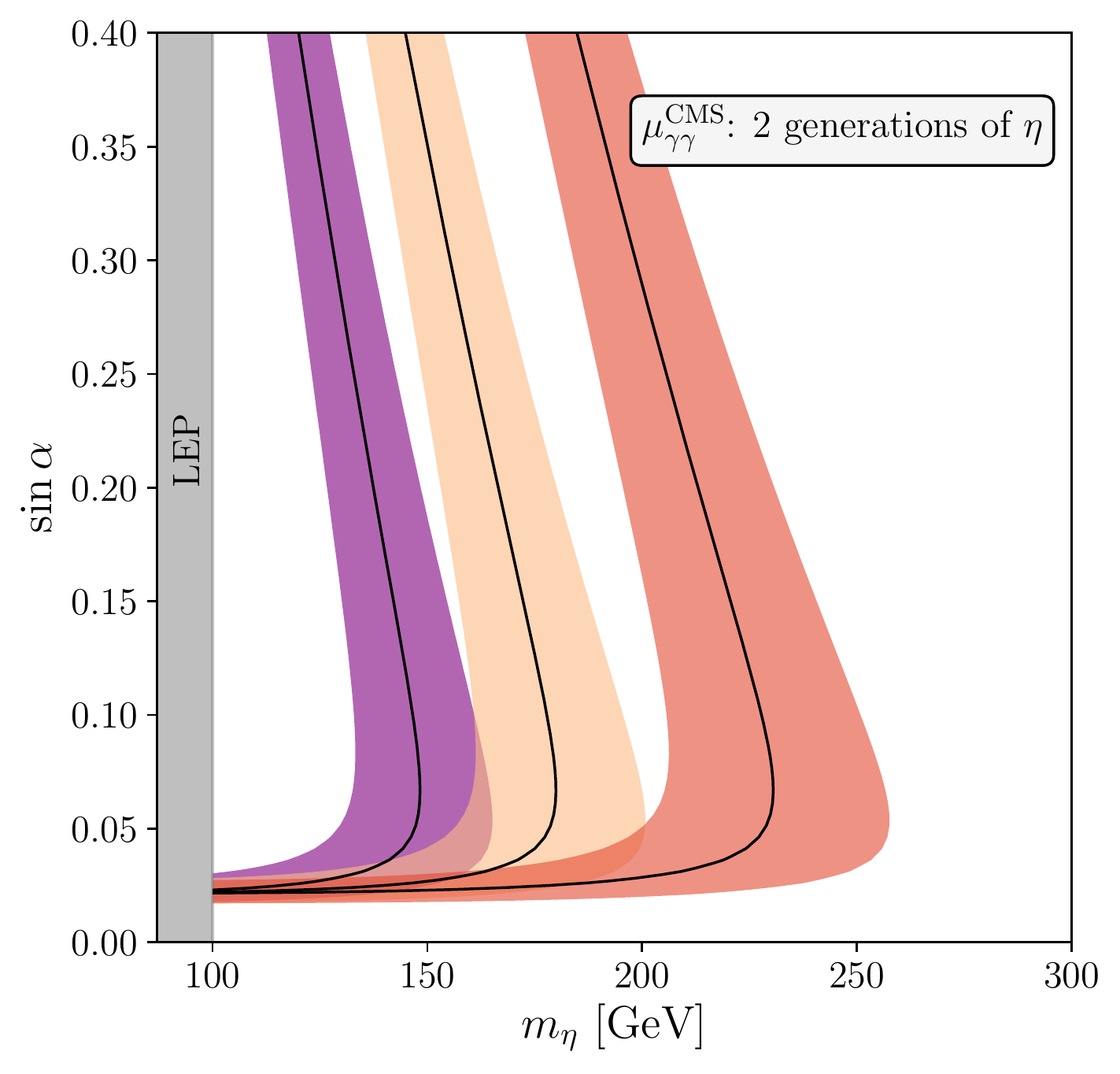} \\
  \includegraphics[width=.43\textwidth]{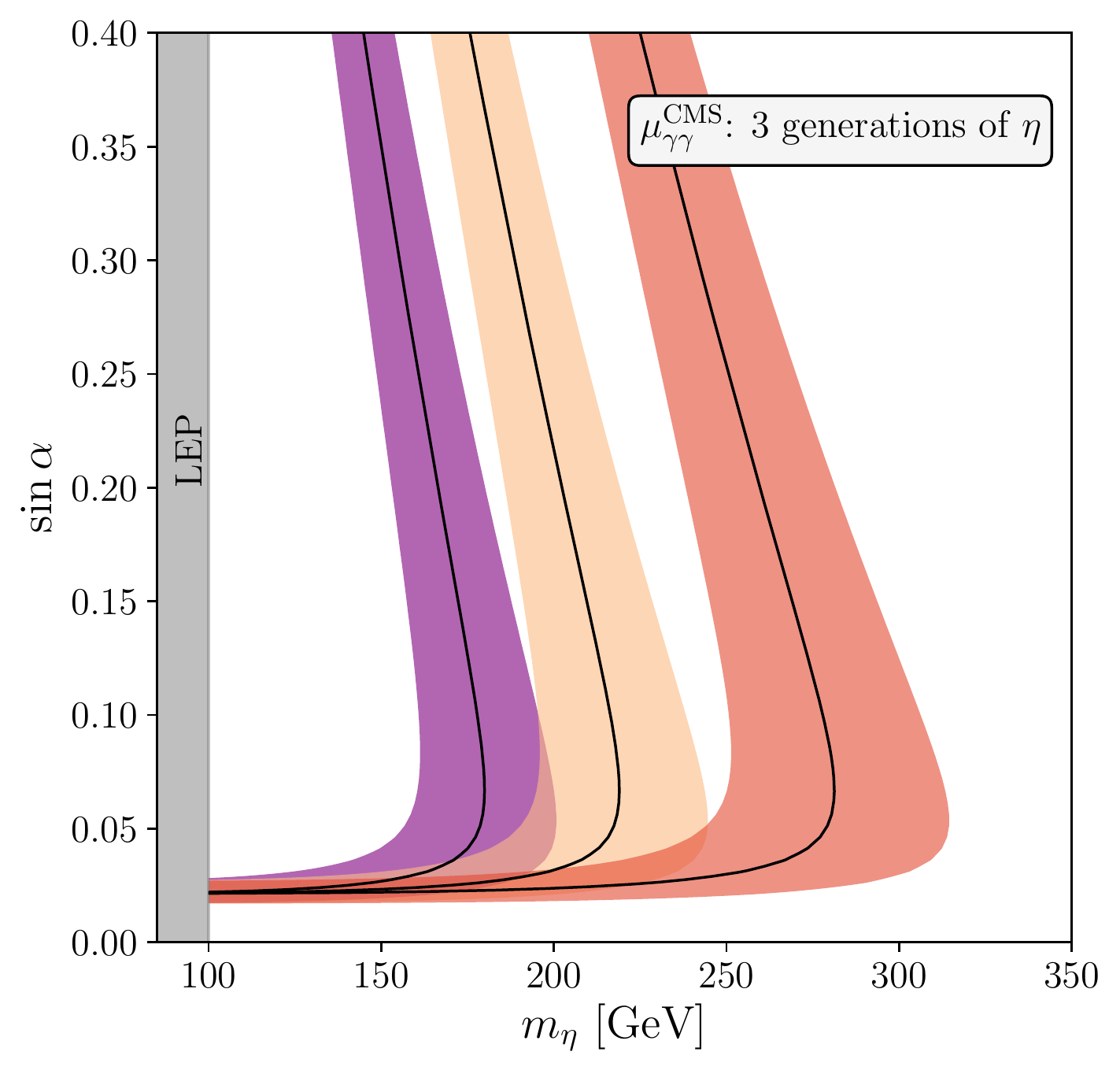}
  \caption{Regions of the $m_{\eta}-\sin\alpha$ plane that
    explain the CMS diphoton excess for fixed values of $\overline{\lambda}_{\eta S} \, v_S = 500$ GeV and $\overline{\lambda}_3 = 0.6$. The colored regions
    accommodate $\mu^{\rm CMS}_{\gamma\gamma}$ at $1 \, \sigma$, 
    while the solid lines correspond to the CMS
    central value $\mu^{\rm CMS}_{\gamma\gamma} = 0.33$. The
    different colors are associated to different values of
    $\overline{\mu}_{\eta}=$ 500 GeV (purple), 1000 GeV (yellow), 2000 GeV (red). The gray band on the
    left is (in principle) excluded by direct searches at LEP,
    although it would be allowed for sufficiently compressed
    spectra.
    \label{fig:mugamma}}
\end{figure*}

Fig.~\ref{fig:mugamma} displays different examples that prove that our
model can easily fit the CMS diphoton excess. In order to obtain these
solutions we have assumed the coupling of the $\eta$ scalars to the
$h_1$ and $h_2$ scalars to be $\overline{\lambda}_{\eta S} \, v_S = 500$ GeV and $\overline{\lambda}_3 = 0.6$, respectively, and we also fixed the value of $\overline{\mu}_\eta$ in the three figures to 500 GeV (purple regions), 1 TeV (yellow regions), and 2 TeV (red regions). 
Then, we found the $(m_{\eta},\sin\alpha)$ pairs that can reproduce
$\mu_{\gamma\gamma}^{\rm CMS}$. Furthermore, we also vary the number
of $\eta$ generations in the model and assume degenerate $\eta$
  doublets. The upper left plot of
Fig.~\ref{fig:mugamma} represents the solution for only one generation
of $\eta$ doublets. We can see that the $m_\eta$ range in which our
model explains the CMS diphoton excess depends on the values of $\mu_\eta$. For example, for the lowest value of the coupling, $\overline{\mu}_{\eta}=500$ GeV, the mass of the charged $\eta$ scalars must
be around $m_{\eta}\sim 100$ GeV to compensate the low value of the
parameter, whereas for larger values of the coupling the $\eta$ masses
can reach $\sim 180$ GeV. In the upper right plot of
Fig.~\ref{fig:mugamma} we can see the case of 2 generations of $\eta$
that have the same mass $m_{\eta_1}=m_{\eta_2}=m_{\eta}$. With two
generations, the diphoton rate increases and, for this reason, the CMS
diphoton excess is explained for greater values of the $\eta$
mass. Something similar happens in the case of 3 generations, shown in
the lower pannel of Fig.~\ref{fig:mugamma}. In this case, the $m_\eta$
value required to accommodate the CMS excess can be as high as 300 GeV
when the $\mu_{\eta}$ couplings are of the order of 2 TeV.

It is important to note that an explanation for the CMS diphoton
excess can be found for small values of $\sin\alpha$, in the $\sim
0.02$ ballpark. This may be surprising at first, since such low values
of the singlet-doublet mixing angle strongly suppress the production
of $h_1$ at the LHC. However, the existence of the low $\alpha$ region
is due to the abovementioned increase in BR($h_1 \to \gamma \gamma$),
see Fig.~\ref{fig:h1BRs}, which compensates for the reduction in the
production cross section. In fact, one can estimate a lower limit on
$\sin\alpha$, below which $h_1$ cannot fit the CMS diphoton signal
strength because the required branching ratio into $\gamma\gamma$
would be larger than 1. The production cross section for a SM Higgs
that decays into a pair of photons at $\sqrt{s}=$ 13 TeV is
approximately $\sigma_{H}\times \BR (H\to \gamma\gamma)\sim 0.125\,
{\rm pb}$~\cite{CMS:2023yay}. Then, assuming the hypothetical scenario
with $\BR (h_1 \to\gamma\gamma)\to 1$\footnote{It is important to note
that this limit is just hypothetical. Once it is reached then the
singlet cannot be produced in the LHC.} one finds the limit
$\sin\alpha \gtrsim 0.0215$ for the central value of $\mu^{\rm
  CMS}_{\gamma\gamma}$ and the range $\sin\alpha \gtrsim
[0.027-0.017]$ taking the $1 \, \sigma$ region. This is precisely what
determines the low $\alpha$ region observed in Fig.~\ref{fig:mugamma}.

\begin{figure*}
  \centering
  \includegraphics[width=.45\textwidth]{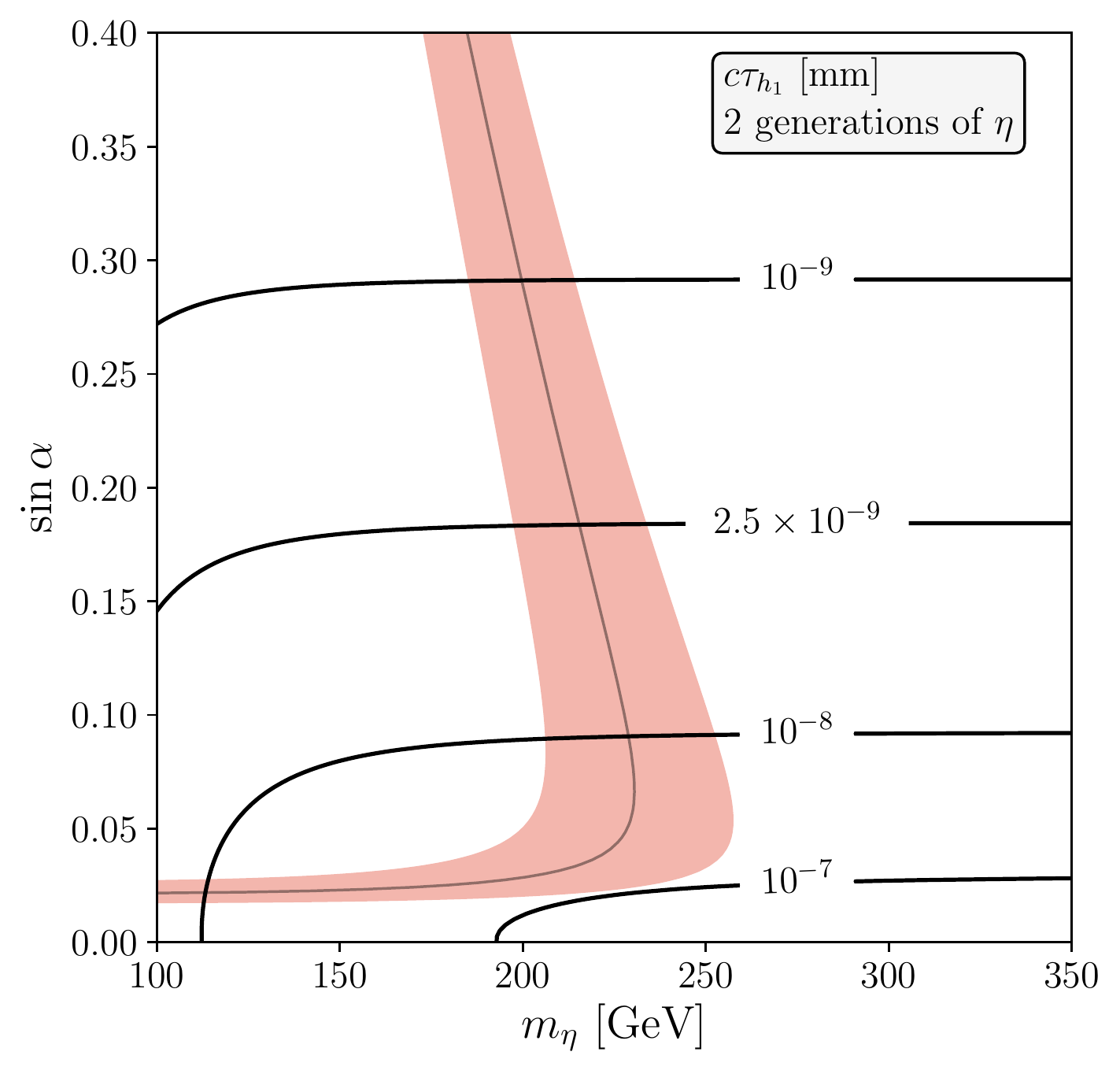}
  \caption{Contours of $\Gamma_{h_1}$ (in GeV) in the $m_\eta - \sin
    \alpha$ plane for a variant of the model with 2 generations of
    $\eta$ doublets. The colored region explains the CMS diphoton
    excess at $1 \, \sigma$ with 2 generations of $\eta$, $\overline{\lambda}_{\eta S}v_s=500$ GeV, $\overline{\mu}_{\eta}=2000$ GeV and $\overline{\lambda}_3 = 0.6$.
  \label{fig:gammah1}}
\end{figure*}

Given that some regions of the $m_{\eta}-\sin\alpha$ plane considered
in Fig.~\ref{fig:mugamma} have small $\sin\alpha$ values, one may
wonder about the decay width of $h_1$. We note that $h_1$ must decay
promptly for our explanation of the CMS diphoton excess to work. We
explore this in Fig.~\ref{fig:gammah1}, which shows contours of $c\tau_{h_1}$ in the $m_\eta - \sin \alpha$ plane for a variant of our model featuring 2 generations of
$\eta$ doublets. We see that $h_1$ has a short decay length, well
below that regarded as prompt, even for small $\alpha$ angles. Again,
the reason is the enhancement of the diphoton decay
channel.

\begin{figure*}
  \centering
    \includegraphics[width=.43\textwidth]{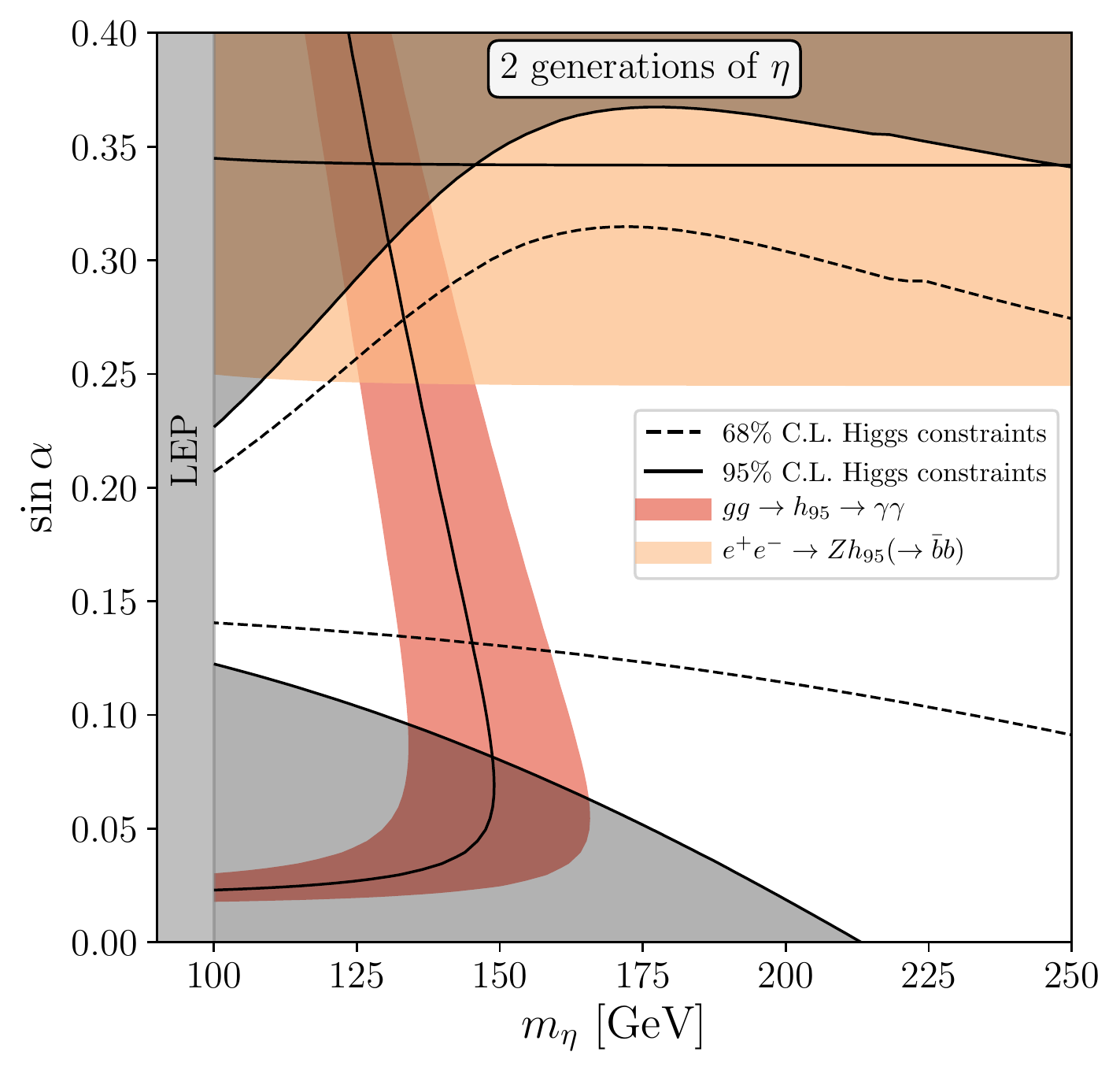}\hspace{1cm}
  \includegraphics[width=.43\textwidth]{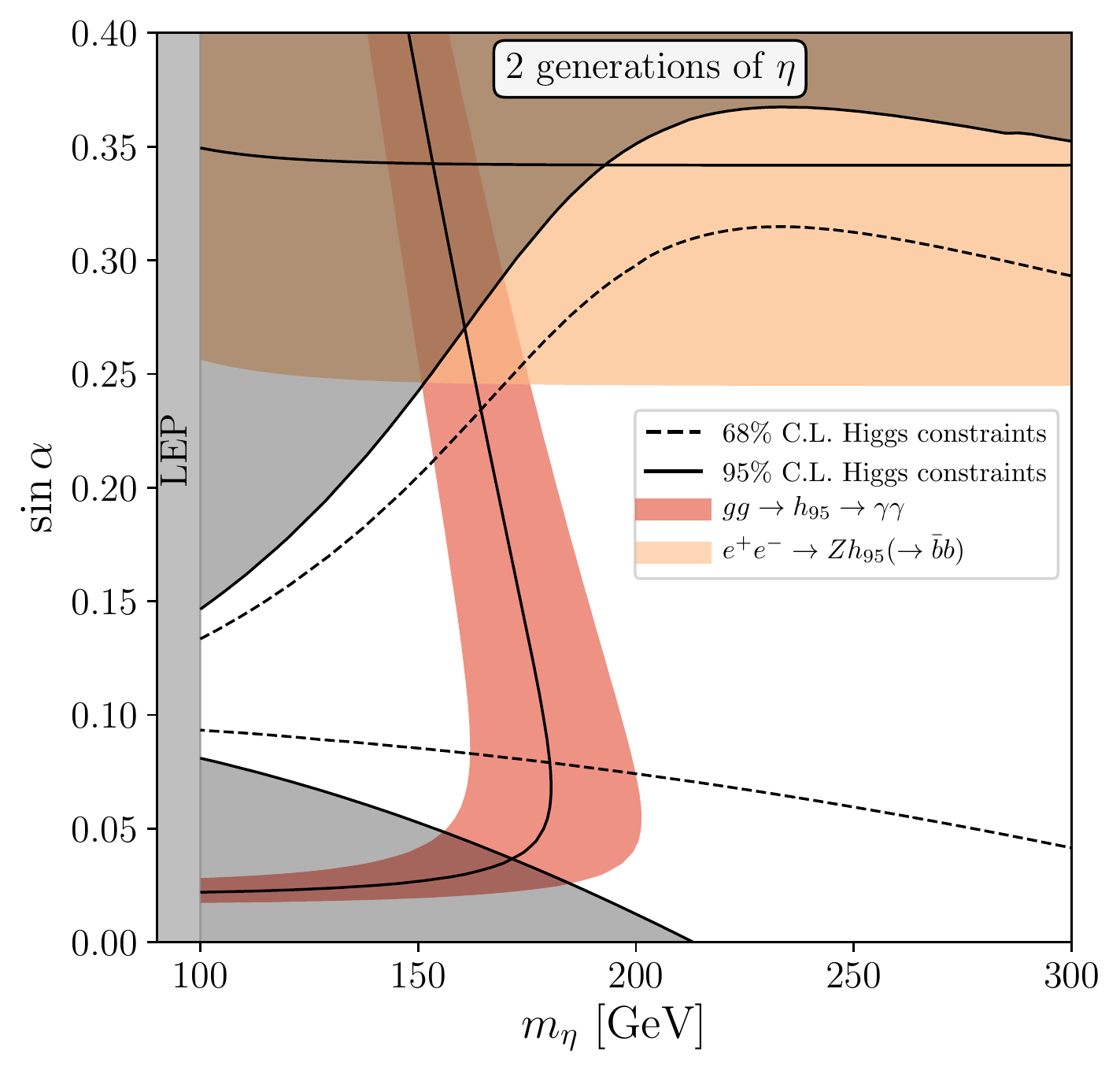} \\
  \includegraphics[width=.43\textwidth]{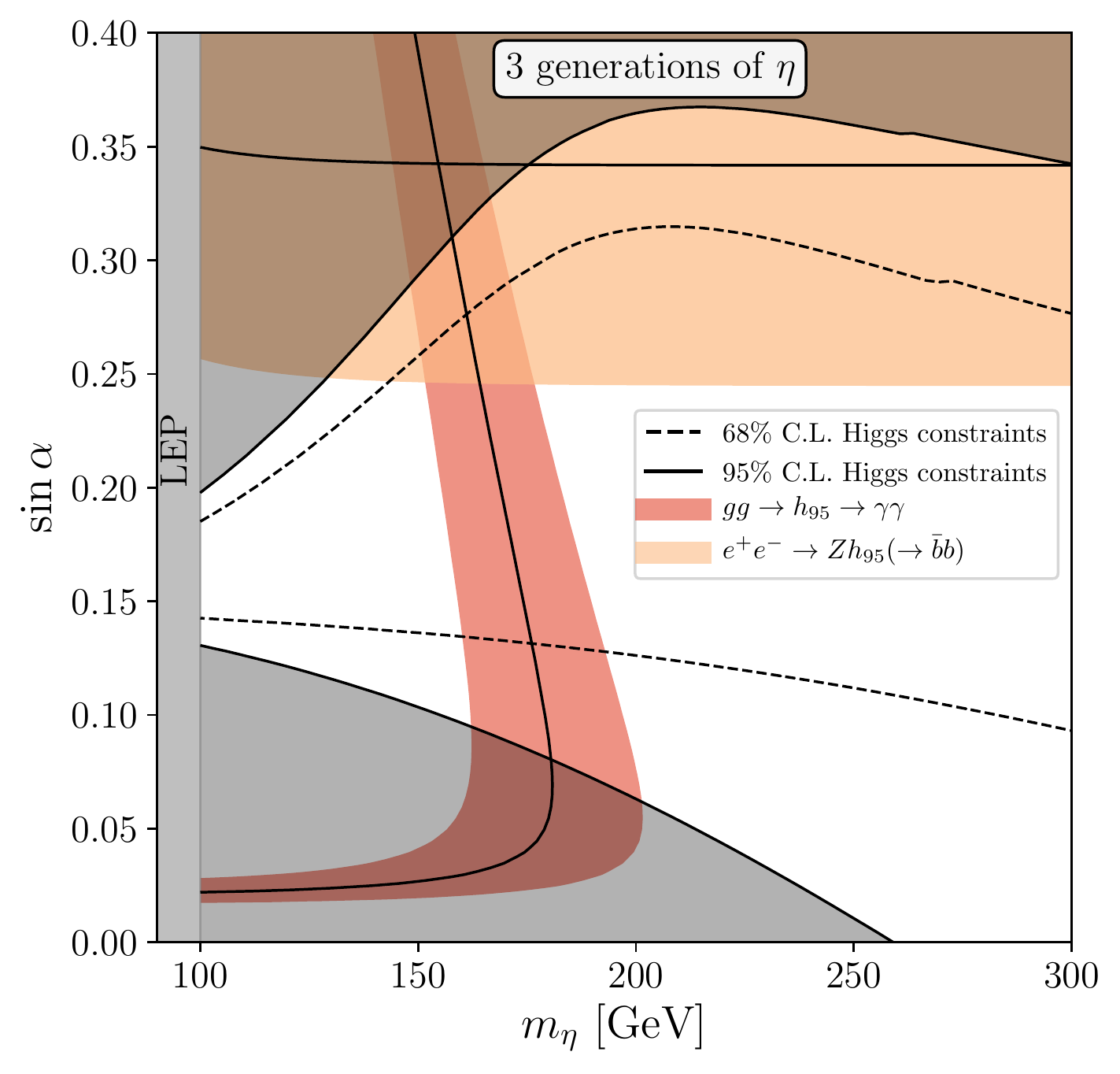}\hspace{1cm}
  \includegraphics[width=.43\textwidth]{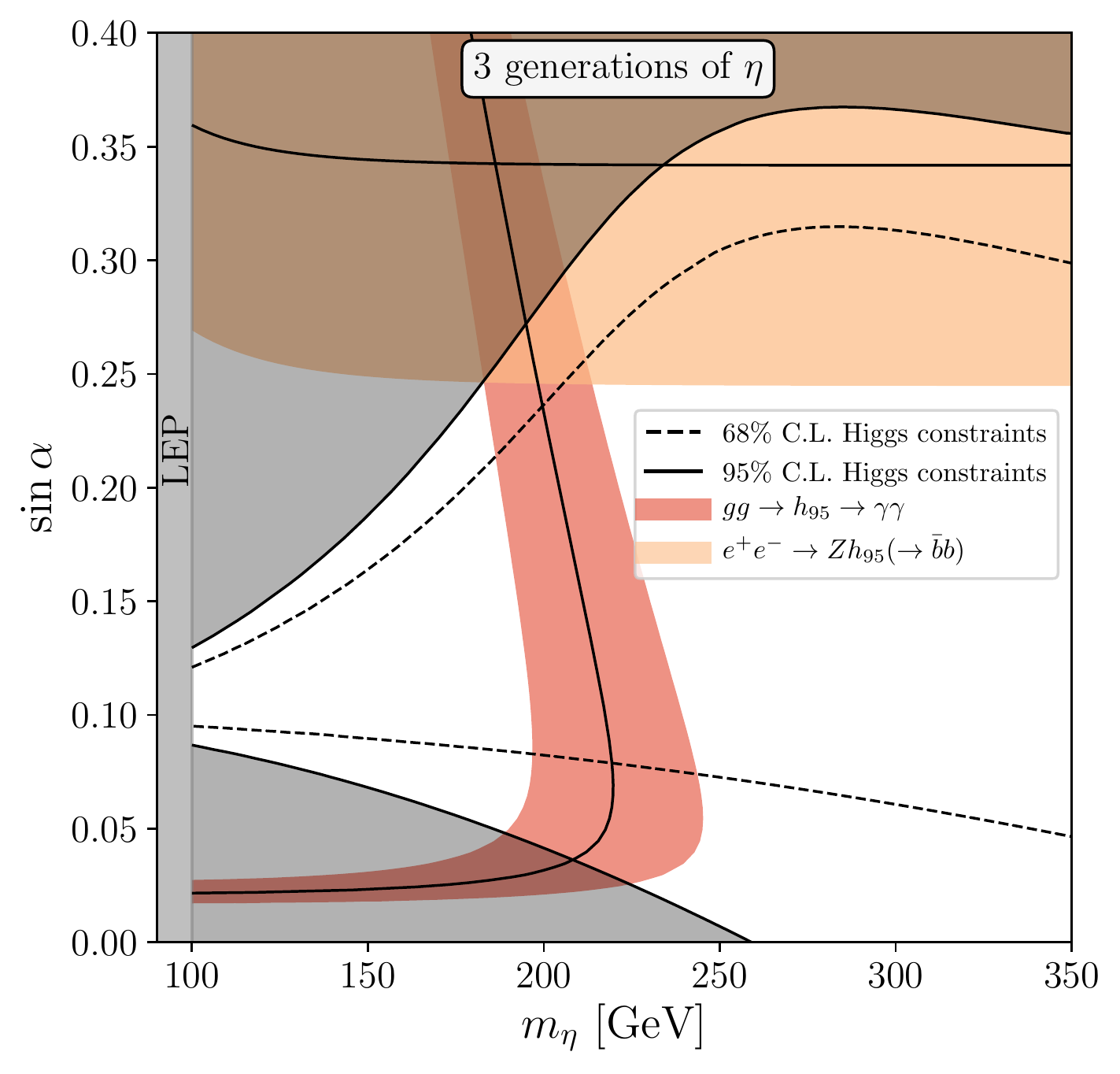}
  \caption{Regions of the $m_{\eta}-\sin\alpha$ plane that explain the
    CMS diphoton and LEP $b \bar b$ excesses. The region that explains
    the CMS diphoton excess at $1 \, \sigma$ is shown in light red for
    $n_{\eta}=2$ (top panels) and $3$ (bottom panels). We have fixed
    $\overline{\lambda}_{S\eta} \, v_S$= 500 GeV and $\overline{\lambda}_3=0.6$, while the
    trilinear $\overline{\mu}_\eta$ takes the values 500 GeV (left panels) and
    1000 GeV (right panels). The area that fits the LEP excess in
    $b\bar{b}$ is shown as light orange while we show as a dark gray
    shaded contour the area that is disfavoured at $95\%$ C.L. by
    Higgs constraints.
  \label{fig:ggvsbb}}
\end{figure*}

Let us now consider the LEP $b\bar{b}$ excess in combination with the
previously discussed CMS diphoton excess. One can see in
Fig.~\ref{fig:ggvsbb} the region of the $(m_\eta,\sin\alpha)$ plane
where both excesses can be explained. As discussed in
Sec.~\ref{sec:inter}, the charged $\eta$ scalars not only affect the
$h_1$ diphoton rate, but also modify the one for $h_2$, already\
measured at the LHC. This implies limits from Higgs data, displayed in
this figure by the dark gray area, which is excluded at $95\%$
C.L.. One should notice that the LEP excess can be explained in a
region of parameter space that lies on a high value of $\sin\alpha$
and is mostly excluded by Higgs data. However, there is still a
portion of the allowed parameter space where both excesses are
explained. As expected, this portion involves lighter charged $\eta$
scalars when fewer generations are considered. With the specific
values chosen in this figure for the $\overline{\lambda}_{S\eta} \, v_S$,
$\overline{\lambda}_3$ and $\overline{\mu}_\eta$ parameters, the required $\eta$ masses
range from $\sim 130$ GeV to $\sim 220$ GeV. Finally, we note that the
dark gray area in Fig.~\ref{fig:ggvsbb} depends very strongly on the
$\overline{\lambda}_3$ value. This can be easily understood by inspecting
Eq.~\eqref{eq:ghetaeta2}. For some values of $\overline{\lambda}_3$, the $g_{h_2
  \eta \eta}$ coupling becomes $\mathcal{O}(1)$ and excludes most of
the parameter space due to Higgs data. However, one can choose
specific values of $\overline{\lambda}_3$ that induce a cancellation in the
$g_{h_2 \eta \eta}$ coupling and make the constraints from Higgs data
less stringent. Our choice $\overline{\lambda}_3=0.6$ is an example of
this.

Alternatively, since the LEP $b\bar{b}$ excess is not very significant
from a statistical point of view, one can interpret it as an upper
limit on the $e^+e^-\to Z \,h_{95} \to Zb\bar{b}$ cross section. In this
case, we conclude that the restrictions imposed by the LEP search in
this channel are compatible with the areas where our model can explain
the diphoton excess in CMS. The case of the CMS ditau excess is
similar. The value of $\sin\alpha$ that would be required to explain
this excess is quite large, above the current limit. The minimum value of the mixing angle in order to explain this signal would be $\sin\alpha_{\rm{min}}\sim 0.7$. However, as there
is only one search by CMS at still low luminosity and there are no
more searches, we consider this excess as still not significant. We
can again interpret it as an upper bound on the $pp\to\,h_{95} \to
\tau^+\tau^-$ cross section. In this case, we conclude again that the
region of parameter space where the CMS diphoton excess is explained
respects this bound.

%New plots:
%\begin{figure*}
%  \centering
%  \includegraphics[width=.43\textwidth]{figs/mu_CMS_vs_bb.pdf} \hspace{1cm}
%  \includegraphics[width=.43\textwidth]{figs/mu_CMS_gg_const.pdf}
%  \includegraphics[width=.43\textwidth]{figs/mu_CMS_gg_2_const.pdf} \hspace{1cm}
%  \includegraphics[width=.43\textwidth]{figs/mu_CMS_gg_3_const.pdf}
%  \caption{ }
%\end{figure*}

\section{Discussion}
\label{sec:discussion}

Once shown that our scenario can accommodate the 95 GeV anomalies, let
us comment on some other aspects of the model that have been ignored
in our previous discussion. This is the case of neutrino oscillation
data. As already explained in Sec.~\ref{subsec:numass}, neutrinos
acquire non-zero masses via loops involving the \z2-odd states $N_n$
and $\eta_a$. Therefore, the resulting neutrino mass matrix depends on
their masses, as well as on the $\lambda_5$ quartics and the $y$
Yukawa couplings~\cite{Escribano:2020iqq}. For specific values of
$n_N$ and $n_\eta$, the $y$ Yukawa couplings can be readily written in
terms of the parameters measured in oscillation experiments using a
Casas-Ibarra parametrization~\cite{Casas:2001sr} adapted to the
Scotogenic model~\cite{Toma:2013zsa, Cordero-Carrion:2018xre,
  Cordero-Carrion:2019qtu}. However, we note that the $y$ Yukawa
couplings do not play any role in the 95 GeV collider phenomenology.

The Scotogenic variant discussed here also contains a DM candidate. In
this family of models one usually has two options: fermion ($N_1$) or
scalar ($\eta_1^0$) DM. However, in order to accommodate the 95 GeV
anomalies we require relatively light $\eta$ doublets, with masses
$m_\eta \lesssim 300$ GeV for $n_\eta \leq 3$.~\footnote{The
explanation of the diphoton excess involves the charged $\eta$ states,
not the neutral ones considered here. However, the mass splitting
between charged and neutral components is small, since it is
controlled by electroweak symmetry breaking effects.} This is too
light to accommodate the DM relic density determined by the Planck
collaboration~\cite{Planck:2018vyg}. In fact, the scalar DM scenario
resembles the Inert Doublet Model~\cite{Deshpande:1977rw}, which is
known to fully account for the observed relic density for DM masses in
the $500-700$ GeV
range~\cite{Barbieri:2006dq,LopezHonorez:2006gr,LopezHonorez:2010eeh,Diaz:2015pyv}. In
contrast, $m_\eta \lesssim 300$ GeV leads to underabundant DM, hence
requiring an additional DM component. We should also note that we
consider more than one generation of $\eta$ doublets. This variation
of the Scotogenic model deserves further investigation, since it may
lead to novel possibilities in scenarios with scalar
DM. Alternatively, we may consider scenarios with fermion DM. This
candidate is known to be potentially problematic due to existing
tension between the DM relic density (which requires large $y$
Yukawas) and contraints from lepton flavor violating observables
(which require small $y$ Yukawas), see for
instance~\cite{Vicente:2014wga}. Two interesting scenarios emerge:
\begin{itemize}
\item $m_{N_1} \ll m_{\eta_1^0}$. If the DM particle $N_1$ is much
  lighter than the $\eta$ states (for instance, $m_{N_1} \sim 100$ GeV
  and $m_{\eta_1} \sim 300$ GeV), the $y$ Yukawa parameters must be
  fine-tuned to suppress the contributions to $\mu-e$ flavor violating
  processes, such as $\mu \to e \gamma$, while being compatible with
  neutrino oscillation data. Although tuned, this scenario is
  possible. It would be characterized at the LHC by the pair
  production (due to the \z2 symmetry) of $\eta$ states which
  subsequently decay into the invisible $N_1$ and leptons: $\eta_1^0
  \to N_1 \, \nu$ and $\eta_1^\pm \to N_1 \, \ell^\pm$. This scenario
  is constrained by existing searches for sleptons, which would have a
  very similar phenomenology in both R-parity conserving and violating supersymmetry
  (see for instance \cite{Dercks:2017lfq, Dreiner:2020lbz, Arganda:2018hdn, Arganda:2021qgi}). Furthermore, a light $N_1$ may also contribute to
  the invisible decay of $h_1$ (if $m_{N_1} \leq m_{h_1}/2$) and/or
  $h_2$ (if $m_{N_1} \leq m_{h_2}/2$). In fact, the $h_1 \to N_1 N_1$
  invisible channel may easily dominate the $h_1$ decay width and
  preclude an explanation of the 95 GeV excess.
\item $m_{N_1} \lesssim m_{\eta_1^0}$. If the $N_1$ singlet is almost
  degenerate with the lightest $\eta$ states, coannihilations become
  efficient and the DM relic density is more easily obtained. This
  enlarges the viable parameter space of the model and leads to novel
  signatures at the LHC. If the mass splitting $\Delta m =
  m_{\eta_1^\pm} - m_{N_1}$ is small enough, the decay $\eta_1^\pm \to
  N_1 \, \ell^\pm$ may involve a long decay length, hence producing
  charged tracks at the detector.
\end{itemize}

Finally, in parameter points with $m_{N_1}\approx m_{h_i}/2$ the annihilation cross section in the early universe gets enhanced due to resonant effects. In such cases one can achieve the correct DM relic density without invoking large couplings. This is a generic feature that does not affect the previous discussion.

\section{Summary and conclusion}
\label{sec:summary}

We have shown that a theoretically well-motivated and economical model
can accommodate the diphoton excess hinted by CMS and ATLAS at 95
GeV as well as the hint for a $b \bar b$ excess at similar energies by
LEP. Our model is a minimal extension of the Scotogenic model and,
besides addressing these collider anomalies, also provides a mechanism
for the generation of neutrino masses and a testable dark matter
candidate. We have allowed for variable numbers of generations of the
Scotogenic states $N$ and $\eta$ and discussed our results for several
choices of interest.

Two CP-even (and \z2-even) scalars: $h_1$ and $h_2$. The lightest of
these states, $h_1$, is identified with $h_{95}$, the hypothetical
scalar that is responsible for the $\gamma\gamma$ and $b \bar b$
excesses at 95 GeV, while $h_2$ is the Higgs-like state discovered by
the CMS and ATLAS collaborations in 2012.  Our numerical analysis
shows that the excesses can be accommodated in our model in a large
fraction of the parameter space. The viable region is characterized by
sizable $\mu_\eta$ trilinear couplings and leads to $\eta$ scalars
with masses below $\sim 300$ GeV ($\sim 180$ GeV) for $n_\eta = 3$
(for $n_\eta = 1$). As expected, a larger number of $\eta$ generations
implies larger contributions to the $h_1$ diphoton coupling and
enlarges the viable parameter space.

Our scenario has a rich phenomenology, both at colliders and at
low-energy experiments. The nature of the dark matter candidate and
the particle spectrum determines the phenomenology at
colliders. Depending on the mass differences between the lightest
\z2-odd fermion and \z2-odd scalar, one expects monolepton events
including missing energy or charged tracks at the LHC detectors. In
addition, the usual lepton flavor violating signatures, common to most
low-energy neutrino mass models, are expected too. Therefore, our
setup not only is well motivated from a theoretical point of view, but
also has interesting phenomenological implications.

We conclude with a note of caution. Although the coincidence of
several excesses, hinted by independent experiments at the same
invariant mass, is highly intriguing, their relatively low statistical
significance implies that more data is required to fully assess their
relevance. We eagerly look forward to future updates on the 95 GeV
excesses.

\section*{Acknowledgments}
The authors are thankful to Thomas Biek\"otter and Sven Heinemeyer for
useful discussions about the 95 GeV excesses and {\tt
  HiggsTools}. This work has been supported by the Spanish grants
PID2020-113775GB-I00 (AEI/10.13039/501100011033) and CIPROM/2021/054
(Generalitat Valenciana). VML is funded by grant Mar\'ia Zambrano
UP2021-044 (ZA2021-081) funded by Ministerio de Universidades and
``European Union - Next Generation EU/PRTR''. AV acknowledges
financial support from MINECO through the Ram\'on y Cajal contract
RYC2018-025795-I.

\appendix

\section{Loop functions}
\label{app:loop}

The loop functions involved in the calculation of $\Gamma(h_i\to
\gamma\gamma)$ are given
by~\cite{Djouadi:2005gi,Djouadi:2005gj,Staub:2016dxq}
\begin{align}
A_0 &=-[\tau -f(\tau)]/\tau^2 \, , \nonumber \\
A_{1/2} &= 2[\tau + (\tau-1)f(\tau)]/\tau^2 \, , \\
A_{1} &=-[2\tau^2+3\tau + 3(2\tau -1)f(\tau)]/\tau^2 \, , \nonumber
\end{align}
where the function $f(\tau)$ is defined as
\begin{eqnarray}
f(\tau)=\left\{ \begin{array}{lc}
\arcsin^2 \sqrt{\tau}; &   \tau \leq 1 \\
\\ -\frac{1}{4}\left[\log \frac{1+\sqrt{1-\tau^{-1}}}{1-\sqrt{1-\tau^{-1}}}-i\pi\right]^2; &   \tau > 1 
\end{array}
\right.
\end{eqnarray}
\vfill
\eject

\bibliographystyle{utphys}
\bibliography{95GeV}

%%%%%%%%%%%%%%%%%

\vfill
\eject

\end{document}